\documentclass[12pt]{article}

\usepackage{amsmath,amsfonts,epsfig,color,latexsym,graphicx}

\newcommand{\be}{\begin{equation}}
\newcommand{\ee}{\end{equation}}
\newcommand{\bea}{\begin{eqnarray}}
\newcommand{\eea}{\end{eqnarray}}

\let\newsection=\section
\renewcommand{\section}{\setcounter{equation}{0}\newsection}

\begin{document}

\begin{flushright}
hep-th/0501038\\
BROWN-HET-1435
\end{flushright}
\vskip.5in

\begin{center}

{\LARGE\bf Heisenberg saturation of the Froissart bound from AdS-CFT}
\vskip 1in
\centerline{\Large Kyungsik Kang and Horatiu Nastase}
\vskip .5in

\end{center}
\centerline{\large Brown University}
\centerline{\large Providence, RI, 02912, USA}

\vskip 1in

\begin{abstract}

{\large In a previous paper, we have analyzed high energy QCD from 
AdS-CFT and proved the saturation of the Froissart 
bound (a purely QCD proof of which is still lacking). In this paper
 we describe 
the calculation in more physical terms and map it to QCD language. We 
find a remarkable agreement with
 the 1952 Heisenberg description of the saturation 
(pre-QCD!) 
in terms of shockwave collisions of pion field distributions. 
It provides a direct map between gauge theory physics and the gravitational 
physics on the IR brane of the Randall-Sundrum model. Saturation occurs
through black hole production on the IR brane, which is in QCD production 
of a nonlinear pion field soliton of a Born-Infeld action
in the hadron collision, that decays into 
free pions.
}

\end{abstract}

\newpage

\section{Introduction}

The Froissart bound \cite{frois} is a bound on the behaviour of the 
total cross section at high center of mass energies $s\rightarrow \infty$, 
with saturation of the type 
\be
\sigma_{tot}\sim \frac{A}{M^2} ln^2 \frac{s}{s_0}
\label{froissart}
\ee
where A is a numerical constant and M is the mass of the smallest excitation 
in the theory. In pure Yang-Mills, M would be the mass of the lightest 
glueball, $M_1= \alpha \Lambda_{QCD}^{-1}$ (we could take by definition 
$\alpha=1$, but we will keep it). If the lightest state is an almost Goldstone
boson, like the pion of QCD, then $M=m_\pi$ and $A\leq \pi$, so that 
$A/M^2\leq 60 mb$. 

In QCD, experimentally, one first found the ``soft Pomeron'' behaviour, 
$\sigma_{tot}\sim s^{0.09}$ at large energies \cite{compas} (cited in the 
2001 PDG \cite{pdgold}) $\sqrt{s}\geq 9 GeV$,
which was then argued to be replaced by a statistically better fit for the 
maximal Froissart behaviour (\ref{froissart}) plus a reaction-dependent 
constant term in 
$\sigma_{tot}$ \cite{compete} (cited in the 2004 PDG \cite{pdg}), that  
fits all data above $\sqrt{s}= 5 GeV$, and with $A/M^2=0.32 mb$, far less 
than $60 mb$. Theoretically,
there is no good explanation for the expected saturation of the 
Froissart bound.  \footnote{For an earlier attempt, based on
 l-plane analiticity, see\cite{kni}.}

Paradoxically, Heisenberg has found in 1952 \cite{heis}
a simple physical model that 
saturates the bound, long before the bound was proposed and even before QCD!
It is a simple effective field theory model, but full of physical insight, 
as we shall see. 

In \cite{kntwo} we have used AdS-CFT \cite{malda}
to analyze the high energy behaviour 
of gauge theories in the large s, fixed t regime. We have found that the 
last energy regime corresponds indeed to the maximal 
Froissart behaviour. Our analysis 
was in the large N, large 't Hooft coupling $g_{YM}^2N$ regime, but we 
have shown that at $s\rightarrow \infty$, the corrections due to finite N 
and $g_{YM}$ are negligible. Thus our proof applies to real QCD as well.

In this paper, we will describe the calculation in physical terms, making 
use of the results in \cite{kntwo}, to which we refer the reader for full 
details. While doing this, we will see a remarkable
agreement with Heisenberg's description and learn how the bound is 
saturated in QCD, while also gaining insight into the dual gravity physics. 
We should note that our analysis only applied to the case where $M_1$ is 
lightest. If $m_\pi$ is lightest, there is just a simple order-of-magnitude 
argument that the bound will be saturated, but is not an
 exact proof. This remains to be investigated in 
further work. For a possible modification of the Heisenberg model to take
into account glueballs, see also \cite{dgn}.

We will first describe Heisenberg's calculation, then our AdS-CFT proof and 
then we will compare them.

\section{ Heisenberg model}

\begin{figure}

\begin{center}

\includegraphics{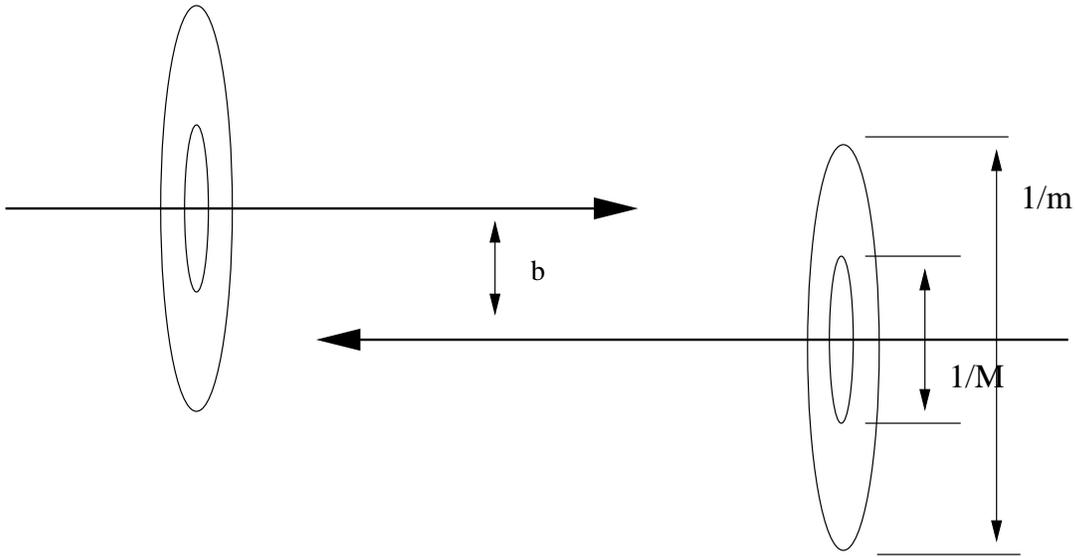}
\end{center}
\caption{Hadron scattering in the center of mass frame. 
M= hadron mass, m= pion mass. Also, A-S shockwave scattering on the IR brane.
M= dual particle size. m=KK graviton mass (gravitational field in 5d, with 
given boundary conditions)}
\label{scatt}
\end{figure}

Heisenberg's
 description starts with scattering of two hadrons of size $\sim 1/M_H$  
in the center of mass frame. Lorentz contraction by the factor $1/\gamma
= \sqrt{1-\beta^2}$ shrinks the size of the hadrons in the direction 
of motion, thus the two colliding hadrons look like pancakes, as in fig.
\ref{scatt}. That is not surprinsing, and one would say that if the 
impact parameter is $b> 1/M_H$, there would be no interaction.

Heisenberg says however that surrounding the hadron there is a pion field 
distribution (cloud of virtual pions), with radius $\sim 1/m_\pi$, also 
Lorentz contracted in the direction of motion, thus also looking like a 
pancake. And in the limit of $s\rightarrow \infty$, when the hadrons and 
the pion distributions will look like shockwaves (zero size in the direction 
of motion, thus delta function distributed), he argues that the hadron size 
becomes irrelevant (we could say that the hadrons ``dissolve'' into the pion
field), and one has a collision of shockwaves of pion field 
distributions. Thus the details of the hadron become irrelevant, and only 
its ability to create pions is a relevant factor.
 
Then the collision of the pion field 
shockwaves is analyzed, and the energy radiated 
away in the collision is calculated. For a free massive scalar pion
\be
(\Box -m_\pi^2)\phi=0
\ee
the energy radiated ${\cal E}$ as a function of the single meson energy $E_0$
(or its de Broglie frequency) is found to satisfy 
\be
\frac{d{\cal E}}{dE_0}= A= {\rm const}.
\ee
up to a maximum energy $E_{0,m}= \gamma m_{\pi}$, and then the number 
of emitted pions satisfies  
\be
\frac{dn}{dE_0}= \frac{A}{E_0}
\ee
Since the pion energy has to be larger than $m_{\pi}$, we get
\be
{\cal E}= A(E_{0, m}-m_{\pi});\;\;\; n= A ln \frac{E_{0,m}}{m_{\pi}}
\ee
which implies that the average energy of emitted 
mesons would increase as s (thus 
also $\gamma\simeq \sqrt{s}/M_H$) increases. 
\be
<E_0>\equiv \frac{{\cal E}}{n}\simeq \frac{E_{0,m}}{
ln \frac{E_{0,m}}{m_{\pi}}}= \gamma m_{\pi}\frac{1}{ln \gamma}
\ee

But this is clearly not satisfactory. In the limit of high energy scattering,
we cannot treat the pion as free. It is 
clear that nonlinearities will flatten the linear growth of $<E_0>$ with 
$\sqrt{s}$. Heisenberg then tried to do it with
a simple $\lambda \phi^4$ interaction, but that clearly didn't work, as 
it is now understood this is 
not a high energy correction. Now we know that the pion 
action is a simple $\lambda \phi^4$ theory, but in terms of 
isomultiplet states of an SU(2) matrix
(the linear sigma model) $\Sigma = \sigma +\tau^a \phi ^a$
\be
{\cal L}_L= \frac{1}{4}tr \partial_{\mu} \Sigma \partial^{\mu}\Sigma
+\frac{\mu ^2}{4}tr \Sigma^+ \Sigma -\frac{\lambda}{16}[tr (\Sigma^+
\Sigma)]^2
\ee
in which 
\be
\Sigma = (v+s)U= (v+s)e^{i\tau^a \pi^a/v}
\ee
and at low energies we can integrate out the ``absolute value'',
the field s, of $\Sigma$,
and get the nonlinear sigma model
\be
{\cal L}_{NL}= \frac{v^2}{4} tr \partial_{\mu} U \partial^{\mu} U^+
\ee
that contains derivative interactions for the pions $\pi^a$. 
As this was  before QCD was discovered, 
and before the pion was described as an isomultiplet of an
SU(2) field, Heisenberg took the 
simplest model in terms of a single scalar field, with a 
remarkable  intuition for the physics, as we see now! 

Indeed, he took the Dirac-Born-Infeld (DBI)-like action for the scalar pion
\be
S= l^{-4}\int d^4x \sqrt{1+l^4[(\partial_{\mu} \phi)^2 +m^2\phi^2]}
\label{dbi}
\ee
with l a length scale, and obtained
\be
\frac{d {\cal E}}{dE_0}= \frac{A}{E_0}\Rightarrow \frac{dn}{dE_0}=
\frac{A}{E_0^2}
\ee
for the same energy region, $m_{\pi}< E_0 <E_{0,m}= \gamma m_{\pi}$, and
\bea
&&{\cal E}= A\; ln \frac{E_{0,m}}{m_{\pi}};\;\;\; n= \frac{A}{m_{\pi}}
(1-\frac{m_{\pi}}{E_{0,m}})\nonumber\\
&&
\Rightarrow <E_0>\equiv \frac{{\cal E}}{n}= m_{\pi} \frac{ln E_{0,m}/m_{\pi}}
{1-m_{\pi}/E_{0,m}}= m_{\pi} \frac{ln \gamma}{1-1/\gamma}\simeq m_{\pi}
ln \gamma
\eea
and now the average energy of the emitted meson is almost independent of 
$\gamma= \sqrt{s}/M_H$, and almost equal to $m_{\pi}$. 

Finally, the last step in computing the cross  section for the pion 
shockwave scattering is to postulate that the energy loss is proportional 
to the total energy, with a proportionality constant that is exponentially
decreasing in the impact parameter 
\be
{\cal E}=\alpha \sqrt{s};\;\;\; \alpha = e^{-bm_{\pi}}
\ee
This is motivated by the fact that the pion distribution has a 
transverse size $\sim 1/m_{\pi}$ and more precisely, the pion wavefunction 
is expected to decrease exponentially with the distance from the hadron, and 
it seems reasonable to assume that the ``degree of inelasticity'' coefficient 
$\alpha$ is proportional to the overlap of pion wavefunctions. 

Then the cross section is found from $\sigma \simeq \pi b_{max}^2$, where 
$b_{max}$ is the maximum impact parameter for which we can still create pions, 
namely when the energy loss ${\cal E}$ is of the order of the average 
emitted pion energy $<E_0>$. Then 
\bea
&&e^{-b_{max}m_{\pi}}\sqrt{s}= <E_0>\Rightarrow b_{max}\simeq \frac{1}{
m_{\pi}}ln \frac{\sqrt{s}}{<E_0>}\nonumber\\
&& \Rightarrow \sigma_{tot}\simeq \frac{\pi}{m_{\pi}^2}ln^2 \frac{\sqrt{s}}{
<E_0>}
\eea
Since $<E_0>$ is almost independent of s, we get the 
maximal Froissart behaviour. 
But note that this behaviour was obtained only because $<E_0>$ was 
independent of s, which came from the DBI-like action for the scalar 
pion. If we had a free pion or a $\lambda \phi^4$ theory, we would not 
get it (we would get a constant $\sigma_{tot}$). One would think
that the minimum 
energy emitted ${\cal E}$ is $m_{\pi}$ anyway, but the correct answer is the 
average pion energy, which for a free pion would grow linearly with $\sqrt{s}$,
and we need the higher derivative, DBI-like action to get $<E_0>\sim m_{\pi}$. 

We should mention here that Heisenberg treats also the case of several 
``pion'' varieties, and finds that the lightest variety dominates the 
high energy behaviour. Also it is clear that the model applies equally well 
to the case of pure gauge theory, when the lightest variety of ``pion'' is 
actually the lightest glueball. This is in fact 
the case that we have studied in detail in AdS-CFT. But we will be a bit 
cavalier in the Heisenberg description and talk about pions and lightest 
glueballs interchangeably.

\section{ Dual saturation of the bound}

Let us now describe our gravity dual (AdS-CFT) version of the saturation 
in \cite{kntwo}. Polchinski and Strassler \cite{ps} (see also \cite{pstwo})
have shown that at 
high energies in a gauge theory, one can describe the scattering of 
colourless states by scattering in a very simple model of gravity dual. 
One takes the $AdS_5\times X_5$ gravity dual to a conformal theory
\be
ds^2= \frac{r^2}{R^2} d\vec{x}^2 + \frac{R^2}{r^2}dr^2 + R^2 ds_X^2= e^{-2y/R}
d\vec{x}^2 + dy^2 + R^2 ds_X^2
\ee
and cuts the warp factor $e^{-2y/R}$ off in the IR, at a $r_{min}\sim 
R^2 \Lambda_{QCD}$ (equivalently, $y_{max}$), which hides our ignorance 
of what happens in the IR (for the gravity dual at small r), that leads the 
theory to become nonconformal. This simple model is enough to 
obtain many features of the nonconformal gauge theory. 

Scattering in the gauge theory of a mode with momentum p and wavefunction 
$e^{ipx}$ corresponds to scattering in AdS of a mode 
with local momentum $\tilde{p}_{\mu}= (R/r) p_{\mu}$ and wavefunction
$e^{i\tilde{p}x}\psi(r, \Omega)$, with $\Omega$ coordinates on X. Then the 
amplitudes in gauge theory are related to amplitudes in AdS by 
\be
{\cal A}_{gauge}(p)= \int dr d^5 \Omega \sqrt{g} {\cal A}_{string}(\tilde{p})
\prod_i \psi_i
\label{polstra}
\ee
High energy scattering in AdS can be defined relative to the string tension
$\alpha ' = R^2/(g_s N)^{1/2}$, and in the gauge theory by the gauge 
theory string tension $\hat{\alpha} '= \Lambda_{QCD}^{-2}/(g_{YM}^2N)^{1/2}$
(with $g_s= g_{YM}^2$). The two are related by $\sqrt{\alpha '}\tilde{p}
_{string}\leq \sqrt{\hat{\alpha}'}p_{QCD}$. We can see that the AdS scale R
corresponds in gauge theory to $\Lambda_{QCD}^{-1}$. 

Giddings then  noticed that one will start producing black holes when one 
reaches the Planck scale $M_P= g_s^{-1/4} \alpha '^{-1/2}$ in AdS, and 
correspondingly $\hat{M}_P= N^{1/4}\Lambda_{QCD}$ in the gauge theory 
\cite{gid}. 
Since the black hole horizon radius in D dimensions grows with energy as 
$r_H\sim E^{1/(D-3)}$, the simplest model for the cross section for black 
hole formation, a black disk with radius $r_H(E=\sqrt{s})$ (all the collision 
energy is taken as mass of the formed black hole), gives
\be
\sigma\sim \pi r_H^2 \sim E^{\frac{2}{D-3}}
\ee
which means (for a 10 dimensional gravity dual), $\sigma \sim s^{1/7} $. 

When the black hole size $r_H$  reaches the AdS size R, we have 
\be 
E\sim M_P^8\sim M_P^8 R^7 \rightarrow E= E_R= M_P(RM_P)^7
\ee
and the corresponding gauge theory energy scale is $\tilde{E}_{R, QCD} = 
N^2 \Lambda_{QCD}$. This is the maximal behaviour one can have, so it 
should correspond to the Froissart behaviour. How do we see that?

The cut-off AdS is just the 2-brane Randall-Sundrum model \cite{rs}, if we 
cut-off also in the UV (unnecessary, but doesn't change physics). If we 
put a point mass of $m=\sqrt{s}$ on the IR brane, we get for the linearized 
metric perturbation 
\be
h_{00, lin} \sim G_4 \sqrt{s} \frac{e^{-M_1r}}{r};\;\;\; G_4^{-1}= M_P^3 R
\ee
where $M_1= j_{1,1}/R$ is the mass of the lightest KK mode
($j_{1,1}$ is the first zero of the Bessel function $J_1$). If we consider 
the position of the horizon of the formed black hole to be roughly when 
$h_{00, lin}\sim 1$, we get 
\be
r_H\sim \frac{1}{M_1}ln (G_4 \sqrt{s}M_1)\Rightarrow \sigma\sim \pi 
r_H^2 \sim \frac{\pi}{M_1^2} ln^2 (\sqrt{s}G_4 M_1)
\ee
and if $\sigma_{QCD}=\sigma$ we obtain the maximal
Froissart behaviour, with mass
scale given by the lightest KK mode, of the order of $R\leftrightarrow 
\Lambda_{QCD}^{-1}$, and corresponds to the lightest pure gauge theory 
excitation= glueball. Note that now indeed $r_H\sim 1/M_1\sim R$, as 
argued.

The case when there is also an almost Goldston boson 
(like the pion) of mass lighter than that of the lightest glueball can 
also be 
modelled by making the radion of the Randall-Sundrum model (the distance 
between the UV and IR branes) dynamical, and giving it a mass $M_L$ by 
a Goldberger-Wise stabilization \cite{gw} or flux stabilization for 
gravity duals of the  Polchinski-Strassler \cite{psthree} type. 
Then, if $M_L < M_1$, the brane will bend under the mass $m=\sqrt{s}$ on the 
IR brane, and the linearized radion change (bending) wil be 
\be
\frac{\delta L}{L}|_{lin} \sim G_4 \sqrt{s} (M_L R) \frac{e^{-M_1r}}{r}
\label{goldst}
\ee
and the maximal Froissart behaviour in gauge theory 
is obtained when the bending in the gravity dual becomes of order 1, 
\be
\frac{\delta L}{L}|_{lin}\sim 1\Rightarrow \sigma_{QCD}= \sigma \sim 
\frac{\pi}{M_L^2}ln^2 (\sqrt{s}G_4 M_L R)
\label{goldstone}
\ee
But there were a number of open questions that we set out to analyze:
Why taking a point mass on the IR brane works? We would like a dynamical 
statement, involving scattering in AdS. Why does $\sigma _{BH}\sim \pi r_H^2$
give a good estimate and why $\sigma = \sigma_{QCD}$? Why does $h_{00, lin}
\sim 1 $ (or $\delta L/L |_{lin}\sim 1$) give a good estimate of the 
horizon (maximum impact parameter) size? And do string corrections, 
corresponding to finite N and $g^2_{YM}N$ corrections in gauge theory, modify 
the results? In answering these questions, we have found a model 
\cite{kntwo} that looks 
remarkably like Heisenberg's.

We want to study the scattering in AdS of two almost
massless modes, at high energy.
The model for this was due to 't Hooft \cite{thooft} and it involves the 
observation that at energies close to the Planck scale (but not above it), 
the massless particles are described  by geometry and gravity alone.
One particle produces a gravitational shockwave, specifically the 
Aichelburg-Sexl \cite{as} solution for flat 4d, of the type
\be
ds^2 = 2dx^+ dx^- +(dx^+)^2 \Phi (x^i) \delta (x^+) +d\vec{x}^2
\ee
The function $\Phi $ satisfies the Poisson equation
\be
\Delta_{D-2} \Phi (x^i) = -16 \pi G p \delta ^{D-2}(x^i)
\label{poisson}
\ee
The second particle is just a null geodesic scattering in this metric, and 
't Hooft showed that this scattering matches Rutherford scattering due to 
single graviton exchange. He also suggested that at energies above $M_{Pl}$ 
both particles should be represented by shockwaves. The reason why only 
gravitational interactions are relevant is because interactions due to massive 
fields are finite range, and at small $r=\sqrt{x^ix^i}$ the function $\Phi$
diverges, thus creating an infinite time delay for the massive interactions. 

In \cite{nastase} the question of putting A-S type shockwaves inside 
a general warped compactication and other manifolds was analyzed. In all 
cases of interest, one adds a shockwave term to the background metric of 
the type $ (dx^+)^2 \Phi (x^i) \delta (x^+)$, where $\Phi$ still satisfies 
the Poisson equation (\ref{poisson}) in the background, and where 
$\Delta_{D-2}$ is the laplacean in the background, at $\partial_{x^+}= 
\partial_{x^-}=0$ (independent of lightcone coordinates). Note that therefore
this is not just the laplacean for an AdS space of lower dimension, since it 
contains the dimension explicitly (as a term $d\partial_y$). 

By comparison, the procedure of calculating the linearized metric 
perturbation $h_{00, lin}$ due to a static point mass, uses the Poisson 
equation with $\Delta_{D-1}$, the laplacean at $\partial_t=0$ (static 
solution), but then there is no a priori reason to expect that the 
full black hole solution still has the same features. By contrast, for 
A-S shockwaves in the spaces of interest (warped compactifications and 
cut-off AdS), we saw that the linearized solution is the exact solution!
For an early attempt at using shockwaves in AdS (similar to ours) for 
AdS-CFT, see \cite{hi}, and later \cite{bf} used shockwave arguments to
argue for the cross section for black hole creation. The linearization 
phenomenon for shockwaves in AdS and on braneworlds was observed in 
\cite{cg}. 

So we have two A-S shockwaves in the gravity dual background scattering,
one travelling in the $x^+$ direction, one in the $x^-$ direction, at
impact parameter b,
but they must create a black hole, which is a highly nonlinear, uncalculable
process. Fortunately, following earlier calculations in flat d=4 in \cite{eg},
generalized by us to curved higher dimensional space in \cite{kn},
we can find when a ``trapped surface'' forms at the interaction point 
$x^+=x^-=0$, and by a GR theorem we know that there will be a horizon forming
outside it, thus a black hole, for which we have a lower bound on the mass.

The last piece of information needed is to turn this classical scattering 
process into a quantum amplitude that we can put into the Polchinski-Strassler
formalism. For 't Hooft scattering, one calculated an amplitude for the 
scattering using an eikonal formalism, finding that $S=e^{i\delta}$ (where 
$\delta $ is the eikonal) is = $e^{i p^+ \Phi}$, with $p^+$ the momentum of 
the second photon, interacting with the A-S solution. Now, we use also an 
eikonal form for the quantum amplitude, with the eikonal being the simplest 
thing one can have, a black disk:
\be
Re(\delta(b,s) )=0\;\;\; Im (\delta(b,s) )=0,\;b>b_{max}(s);\;\;\;
Im (\delta(b,s) )=\infty , \; b<b_{max}(s)
\ee
where $b_{max}(s)$ is what we find from the classical A-S scattering. Then 
the imaginary part of the forward amplitude reproduces the classical 
$\pi b_{max}^2(s)$ result for $\sigma_{tot}$, but now we have a quantum 
$2\rightarrow 2$ amplitude that we can put in the Polchinski-Strassler formula
(\ref{polstra}). 

When we plug in several relevant forms for $b_{max}(s)$ like $as^{\beta}$ and
$a \; ln s$, we find that when we translate to gauge amplitudes, all we 
get is we multiply $\sigma_{string}$ with a model-dependent constant, and 
we modify the subleading behaviour. But more importantly, we find that 
most of the extra dimension (r)
integration in the gauge theory amplitude is concentrated near the IR brane, 
for these high energy behaviours. 

We have seen that in the simple Giddings description \cite{gid}, Froissart 
behaviour should come in when the black hole size reaches the AdS size.
But if most of the gauge theory amplitude comes from near the IR brane, 
the scattering will look as if it happens on the IR brane itself (due to 
its large size, the formed 
black hole will not ``see'' that is outside the IR brane). The A-S shockwave
solution living on the IR brane is found to be 
\bea
\Phi (r,y)
&=& \frac{4G_{d+1}p}{2\pi}
e^{-\frac{d|y|}{2R}}\int \frac{d^{d-2}\vec{q}}{(2\pi)^{d-2}}e^{i\vec{q}
\vec{x}}\frac{I_{d/2}(e^{-|y|/R}Rq)}{qI_{d/2-1}(Rq)}\nonumber\\
&=&\frac{4G_{d+1} p}
{(2\pi)^{\frac{d-4}{2}}}\frac{e^{-\frac{d|y|}{2R}}}{r^{\frac{d-4
}{2}}}\int_0^{\infty} dq q^{\frac{d-4}{2}}J_{\frac{d-4}{2}}(qr)\frac{I_{d/2}
(e^{-|y|/R}Rq)}{I_{d/2-1}(Rq)}
\eea
This solution is found by imposing normalizability at $y=\pm \infty$ (on the 
UV brane), and matching conditions (periodicity) at the IR brane for the 
Poisson equation solution. Therefore it can be defined as the first KK mode
for the effective theory of massless modes on the IR brane. At large r it 
becomes
\be
\Phi(r, y=0)\simeq R_s \sqrt{\frac{2\pi R}{r}}C_1 e^{-M_1 r}; \;\;
C_1=\frac{j_{1,1}^{-1/2}J_2(j_{1,1})}{a_{1,1}};\;\; J_1(z)
\sim a_{1,1}(z-j_{1,1});\;\;z\rightarrow j_{1,1}
\label{kkwvfct}
\ee
and we can see the same exponential drop as in $h_{00,lin}$, only the power 
or r is different, due to the fact that we have a solution of $\Delta_{D-2}$ 
(massless perturbation), as opposed to $\Delta_{D-1}$ (static massive 
perturbation). 

If we take two A-S waves on the IR brane scattering at b=0, the condition 
that determines the shape and size of the trapped surface is 
\be
(\nabla\Psi)^2+ e^{2|y|/R}(\partial_y \Psi)^2=4;\;\;\; \Psi= \Phi + \zeta
\ee
where $\zeta$ is defined perturbatively in y by the condition that the above
matches also $\Psi=C$=const.(for full details see \cite{kn} and \cite{kntwo}).

One finds that the condition for the trapped surface size r at y=0 is 
\be
\frac{3r}{2R^2}\Phi (r, y=0)=1
\ee
which we see that is similar to the approximate condition for the 
horizon $r_H$ 
that \cite{gid} had, namely $h_{00, lin}\sim 1$ (but the power of r and the 
constants are different). 

But one can do better, one can find an approximate condition for the 
trapped surface at nonzero b, 
\be
(\frac{3r}{2R^2}\Phi (r, y=0))^2 (1-\frac{b^2}{2r^2})=1
\label{bhform}
\ee
which gives a maximum b that satisfies it that is approximately
\be
b_{max}(s)= \frac{\sqrt{2}}{M_1}ln [R_s M_1 K];\;\;\; K= \frac{3\sqrt{\pi}}{
\sqrt{2}j_{1,1}^{3/2}}\simeq 0.501
\ee
and $R_s= G_4\sqrt{s}, G_4= 1/(RM_{P,5}^3), M_1= j_{1,1}/R$ ($j_{1,1}\simeq 
3.83$).

As we mentioned, then the the gauge theory cross section is (via 
the Polchinski-Strassler formalism)
\be
\sigma_{tot}= \bar{K}\pi b_{max}^2(\tilde{s})
\ee
with $\bar{K}$
 a model dependent constant, $\tilde{s}= s\hat{\alpha}'/\alpha '$, or 
equivalently by keeping s fixed and replacing $R$ by $\Lambda_{QCD}^{-1}$
and $M_{P,5}$ by $N^{1/4}\Lambda_{QCD}$ (gauge theory quantities), 
and we get the expected Froissart behaviour.

Up to now we have discussed strictly speaking the usual AdS-CFT limit, 
of large N and large $g^2_{YM}N$, since this corresponds to small $\alpha '$
and $g_s$ string corrections in the gravity dual. But in \cite{kn} we have 
also analyzed string corrections using a model by Amati and Klimcik \cite{ak}.
They obtained a string-corrected A-S wave by matching the 't Hooft scattering 
of a superstring in an arbitrary shockwave profile $\Phi$, $S= e^{ip^+\Phi}$,
with a resummed eikonal superstring calculation $S= e^{i\delta}$
\cite{acv87}, and finding the $\Phi$ that equates the two results. Then
\be
\Phi (y)= - q^v \int_0^{\pi} \frac{4}{s} : a_{tree} (s, y- X^d(\sigma_d, 0): 
\frac{d\sigma_d}{\pi}
\ee
where $2 p_v q^v= -s$ and $b= x^u-x^d$ is the variable y. 
Note that this corresponds both to $\alpha '$ corrections, given by 
$a_{tree}$, and to $g_s$ corrections, since the eikonal form $e^{i\delta}$
resummed ``ladder diagrams'', which are predominant at $s\rightarrow 
\infty$, $e^{i\delta}=\sum_h (i\delta)^h/h!\sim \sum_h (g_s)^h (a_{tree})^h
/h!$ (h= loop number). We have found that 
by scattering two of these modified A-S waves the effect on $b_{max}$, now 
called $B_{max}$ to avoid confusion with $b\sim y$ is 
\be
B_{max}= \frac{R_s}{\sqrt{2}}(1+ e^{-\frac{R_s^2}{8\alpha' log (\alpha's)}})
\ee
in the regime where the exponent is large (in absolute value). Thus at 
$s\rightarrow \infty$, the string corrections to black hole production are 
exponentially small! Although this result was obtained for flat 4d, it is 
not hard to imagine that it will remain true in the warped compactification 
case, for scattering on the IR brane. 
And as string corrections are mapped to large N, large 't Hooft coupling
corrections, we can say that the Froissart behaviour will also apply for the 
case of real QCD (small N, small 't Hooft coupling)!

Note that we are talking here about string corrections to the scattering 
itself, but of course there will be corrections to the gravity dual 
background. Since however our model was so general (no model really, 
just cut off AdS), we can confidently say that all that can happen is 
for the parameters of the theory, the AdS size R and $M_P$, to get 
renormalized. That would translate into gauge theory to a modification of 
the energy scales $\Lambda_{QCD}, \hat{M}_P$ and $\hat{E}_R$, the scale 
of the onset of Froissart behaviour.

\section{ Comparison to QCD and Heisenberg}

So then we can ask the question how does our calculation translate into 
QCD and Heisenberg language? 

We know since shortly after the Randall-Sundrum model was proposed 
\cite{rs,rstwo} that 
we can understand it as just AdS-CFT when gravity is not decoupled from the 
4d physics \cite{verlinde}. As usual in the AdS-CFT correspondence, KK 
modes of the graviton in 5d correspond to glueball excitation of the 
gauge theory. One can understand RG flow of the gauge theory (scale 
transformations) as just motion in the 5th direction in the gravity dual
(as if we move a physical brane in the 5th direction). On the other hand,
bulk gravity can be reduced to usual gravity+ KK modes on the IR brane, 
coupled to the Standard Model that might live there. So on the IR brane,
there is a duality between the gauge physics of glueball states and the 
gravity physics of KK  modes, as both have their origin in bulk gravity. 

With Polchinski-Strassler, we have made concrete the AdS-CFT duality part,
with the gauge theory living on a 4d brane, not necessarily the IR brane. 
Scattering in the gauge brane corresponds to scattering in AdS, which itself 
can be reduced in certain cases to scattering in an effective field theory 
on the IR brane. 

Indeed, we have found that in the Froissart regime most of the AdS scattering 
happens near the IR brane, and we can effectively describe it as scattering 
on the IR brane. The A-S shockwaves have profiles that correspond to KK 
gravitons (solutions to the Poisson equation with certain boundary conditions 
on the UV and IR branes).

From 't Hooft we know that corrections at large s
due to massive modes are negligible, and only gravity is 
relevant. We represent the scattering of dual particles just by scattering 
of gravitational A-S shockwaves. Thus the same fig. \ref{scatt} can be 
used to describe the dual picture as well! We have shockwaves (limits of
pancake-like distributions) colliding, and the particles dual to the hadrons 
``dissolve'' into the KK graviton field. Indeed, as we said, from the 4d 
point of view, the wave profile $\Phi$ corresponds to the first
 KK graviton mode. Moreover, the A-S shockwaves can be actually found 
by boosting black holes to the speed of light (and keeping their energy 
fixed), see \cite{as} and \cite{nastase}, the same way we boost hadrons.  

The first KK graviton mode corresponds to the lightest glueball, which as 
we argued should replace the pion in the Heisenberg analysis for the 
case where the pion is heavier, so we have a direct correspondence. In 
Heisenberg's analysis, the ``degree of inelasticity'' was postulated
to be $e^{-bm_{\pi}}$, based on the fact that the pion wavefunction 
around the hadron should go like $e^{-rm_{\pi}}$.
Now we have a form for the KK graviton wavefunction (\ref{kkwvfct}), 
and a dynamical mechanism to check for the creation of the black 
hole, as given in (\ref{bhform}). 

So what happens in QCD when a black hole forms in the dual description?
Pretty much the same, in a certain sense. As Heisenberg describes, if we 
had a free massive pion, the shockwave collision will produce in the 
interaction regime (after the collision) just free spherical waves
(like two small plane water waves colliding in a pond). This would correspond
on the dual side to having a free graviton. Clearly, nothing would happen 
there, and as Heisenberg described, in that case $<E_0>\sim \sqrt{s}$ and 
thus a constant $\sigma_{tot}$. We need to take instead a nonlinear pion
as in the DBI-like action (\ref{dbi}). Then as we said $<E_0>$ is almost 
independent of s and we find the maximal 
Froissart behaviour. Heisenberg actually 
calculates perturbatively the pion field in the interaction region, but 
it can't be calculated everyhwere. One should find the equivalent of the 
black hole formation for the scalar pion, i.e. a highly singular 
nonlinear structure. For A-S collision in flat 4d, the metric after 
the interaction was also found perturbatively in \cite{dp}, but one 
can't draw any conclusions from it about the nonperturbative solution. 
Luckily, we had the trapped surface formalism that relied on the metric at 
the interaction point. 

So a dual black hole is in QCD a highly nonlinear soliton being formed in the 
collision of two pion field distributions, and that further decays into 
free perturbative waves (radiated pions). Heisenberg couldn't calculate 
the soliton form, but he could calculate the average energy of the pions 
emitted in the decay of the soliton, finding as we saw $<E_0>\simeq 
m_{\pi} ln \gamma$. That, coupled with the assumption of 
``degree of inelasticity'' mirroring the pion field distribution around the 
hadron, was enough to derive the maximal Froissart behaviour. 

In our case, we have the wavefunction profile $\Phi$ in (\ref{kkwvfct}) 
that shows exponential decay, and then we use the full 5d GR rules to 
calculate whether a black hole (pion field soliton) forms and decays, as 
in (\ref{bhform}), so we don't need to calculate the average energy of 
gravitons emitted in the decay of the black hole and postulate a ``degree
of inelasticity'' for the energy loss, but the idea is the same!
Presumably, there should be an effective purely 4d description of the 
KK graviton mode $\Phi$ of mass $M_1$ that plays the role of the DBI 
action (\ref{dbi}) for Heisenberg. It is tempting to assume that it is 
exactly the same action.

In fact, we have been talking about pions until now, but as we mentioned 
we should really speak about lightest glueballs, as our analysis was 
done only for that case. But the simple analysis in \cite{gid} that we 
mentioned before applies also for an almost Goldston boson like the radion 
of Randall-Sundrum (\ref{goldst}),(\ref{goldstone}). The remarkable thing 
is that the action for the massless 
radion (position of the IR brane) is the DBI 
action! Indeed, we can easily check that for a codimension one brane, 
in the straight gauge $X^{\mu}: (X^a=\xi^a, X)$ (where $\xi^a$ are worldvolume
coordinates and $X= l_s^2 \phi$ is the radion= 5th coordinate, and we did 
the rescaling to a canonical dimension 4d field $\phi$), the action is 
\be
{\cal L}= l_s^{-4}\sqrt{det(\partial_a X^{\mu} \partial_b X^{\nu} g_{\mu\nu})}
= l_s^{-4}\sqrt{g}\sqrt{1+(\partial_a X)^2}= 
l_s^{-4}\sqrt{g}\sqrt{1+l_s^4(\partial_a \phi)^2}
\ee
where the contraction in the square root and the metric g on the right hand 
side are done with the 4d reduced metric $g_{ab}$.  
And Heisenberg's suggestion that the pion mass be put inside the square root
as in (\ref{dbi}) is suggestive that maybe one should try to do the same 
in radion stabilization mechanisms. We are not aware whether this was 
considered in the literature, but since as Heisenberg points out this is 
relevant for getting the right nonlinear behaviour in the QCD side, 
one should perhaps consider it as a nonlinear extension of the Goldberger-Wise 
stabilization \cite{gw}.

Also, we have expressed the hope that the KK graviton and black hole 
creation can be described by a DBI-like action, maybe exactly (\ref{dbi}) for 
the wavefunction $\Phi$, but it is not clear that one can. After all, we made 
use of the full GR at least to deduce from the appearence of a trapped surface
that a horizon will form outside it, and probably the whole nonlinear and 
tensor nature of gravity is needed. But one could maybe look for a 
4d effective action for the massive KK graviton $g_{\mu\nu}^{(1)}$. 
Born-Infeld-type actions for gravity have been considered before.
For instance, \cite{dg} analyzes a more general action,  of the type 
\be
{\cal L}= \sqrt{-det(g_{\mu\nu}+ aR_{\mu\nu} +bX_{\mu\nu})}
\ee
where $X_{\mu\nu}$ is an expression that can be quadratic or higher in 
curvatures and can be used to put almost any action in this DBI 
form. More to the point, the action for b=0, a=1 (true Born-Infeld), 
when written in first order form (see  for instance \cite{nast})
\be
\int \sqrt{ det (R^{ab}(\omega) + l^{-2}e^a e^b) }
\ee
can be rewritten as a Lanczos-Lovelock action, is the equivalent of the 
odd dimensional Chern-Simons action for gravity (as a gauge theory of the 
Poincare group), and in fact can be obtained
by dimensional reduction from it, in any dimension \cite{nast}. In 4d, 
it is rewritten as (the contraction of local gauge indices is done with
$\epsilon^{abcd}$)
\be
R\wedge R + 2l^{-2} R\wedge e\wedge e + l^{-4} e\wedge e\wedge e\wedge e
\ee
where the first term is topological, the second and third are Einstein-Hilbert
and cosmological constant terms. Thus this action is the usual Einstein-AdS 
theory with a topological term, so regular 4d gravity is already a Born-Infeld
type action. 

Perhaps also the hoped-for effective action for the KK graviton looks like 
a BI action, something like 
\be
\sqrt{det(g_{\mu\nu} +R_{\mu\nu}(g_{\rho\sigma}^{(1)})+X_{\mu\nu}
(g_{\rho\sigma}^{(1)}))}
\ee
where $g_{\mu\nu}$ is the massless 4d graviton and $g_{\rho\sigma}^{(1)}$ is 
the massive KK graviton, with $R_{\mu\nu}$ bilinear in 
$g_{\rho\sigma}^{(1)}$ (and the rest of metric fields are $g_{\mu\nu}$)
and $X_{\mu\nu}$ a mass term also bilinear in it. 

So we used Heisenberg's description to make some 
conjectures about the dual 
gravity theory (for a radion and KK graviton effective action), and from 
the gravity dual calculation we have found a precise description of 
the mechanism for Froissart saturation. But this mechanism is in terms of 
the effective field theory of pions and lightest glueball states. 
It is clear that this is the only thing that we can 
learn from AdS-CFT, as AdS-CFT deals with gauge invariant quantities. 
However, maybe this precise description can be used to learn about a 
QCD proof of the saturation too. Finally, it would be nice to have a 
precise dual description for the case the pion is the lightest field too.

In conclusion, one can only be amazed by Heisenberg's physical insight,
well ahead of his time. His effective field theory description matches 
exactly the gravity dual description of the saturation. His use of the 
DBI action for the pion describes the radion action and maybe the KK 
graviton. The DBI action seems to be in the same universality class as 
the action for the real (SU(2)) pions, generating the same physics. 

{\bf Acknowledgements}. We would like to thank Micha Berkooz for discussions.
This research was  supported in part by DOE
grant DE-FE0291ER40688-Task A.

\end{document}